\documentclass[twocolumn,showpacs,preprintnumbers,amsmath,amssymb]{revtex4}

\usepackage{graphicx}
\usepackage{dcolumn}
\usepackage{bm}
\usepackage{amsfonts}%
\usepackage{amsmath}

\begin{document}


\title{Multichannel cold collisions between metastable Sr atoms}
\author{Viatcheslav Kokoouline, Robin Santra, and Chris H. Greene}
\affiliation{Department of Physics and JILA, University of Colorado, 
Boulder, Colorado 80309-0440, USA}
\date{\today}

\begin{abstract}
We present a multichannel scattering calculation of elastic and 
inelastic cold collisions between two low-field seeking, metastable $^{88}$Sr 
[$(5s5p) ^3P_2$] atoms in the presence of an external magnetic field. The 
scattering physics is governed by strong anisotropic long-range interactions,
which lead to pronounced coupling among the partial waves of relative motion.
As a result, nonadiabatic transitions are shown to trigger a high rate of 
inelastic losses. At 
relatively high energies, $T > 100\mu $K, the total inelastic collision rate
is comparable with the elastic rate. However, at lower collisional energy, the 
elastic rate decreases, and at $T\sim 1\mu$K, it becomes substantially smaller
than the inelastic rate. Our study suggests that magnetic trapping and 
evaporative cooling of $^{88}$Sr [$(5s5p) ^3P_2$] atoms, as well as $^{40}$Ca  
[$(4s4p) ^3P_2$], in low-field seeking states will prove difficult to achieve
experimentally.
\end{abstract}

\vspace*{0cm}

\maketitle

In recent years, the trapping and cooling of alkaline-earth atoms has been pursued with increased interest. Trapping experiments with $^{88}$Sr \cite{DiVo99,katori99,katori01,nagel03,xu03} and $^{40}$Ca \cite{zinner00,HaMo03} atoms have been carried out, and experiments are underway in a number of laboratories. One major thrust of this research track is to extend the small class of atomic species that have been Bose-Einstein condensed. Also, the most prevalent isotopes have closed-shell nuclei, which eliminates complications from hyperfine structure and simplifies the interpretation of cold collision data \cite{MaJu01}. 

The usual route to quantum degeneracy utilizes magnetic trapping and evaporative cooling of precooled atoms. Since ground-state alkaline-earth atoms possess no magnetic dipole moment, some experimental groups work with a system of alkaline-earth atoms in their lowest $(nsnp) ^3P_2$ metastable state \cite{LoBo01,katori01,nagel03,xu03,HaMo03}. The radiative lifetime of this state is of the order of ten minutes \cite{Dere01}. However, as we show in this Letter, it is not spontaneous emission that ultimately limits the magnetic-trap lifetime of metastable alkaline-earth atoms. Their nonspherical electronic structure gives rise to strong anisotropic interatomic interactions not present in ground-state alkali metals, hydrogen, or metastable helium. In the present study, these interactions are predicted to generate severe inelastic losses.

Derevianko {\em et al.} \cite{derevianko03} were the first to theoretically approach the intriguing problem of anisotropic long-range interactions between metastable alkaline-earth atoms in an external magnetic field. Their study revealed important insights, notably the existence of a long-range molecular potential well. They also provided a first estimate of inelastic losses within a two-channel model, but at a qualitative level only. Recently, Ref. \cite{santra03} performed a detailed tensorial analysis of the anisotropic long-range forces between two nonspherical atoms in a magnetic field  and investigated the associated coupling between electronic structure and molecular rotational motion. A two-body potential was derived that is ideal for multichannel-scattering calculations. The present work has the goal to carry the aforementioned theoretical studies to the next level, and thus to provide guidance to experimentalists. Specifically, we develop a quantitative quantum description of the two-body elastic and inelastic collisions between alkaline-earth atoms in the $(nsnp) ^3P_2$ excited state. We focus on strontium atoms but briefly address implications of the present study to Ca and other similar atoms. 

Our treatment is based on the multichannel potential derived in Ref.~\cite{santra03}:
\begin{eqnarray}
\label{eq:Hele}
\hat V & = & \frac{{\bm L}^2}{2 m_{\mathrm{red}} R^{2}} + \hat H_{\mathrm{I}}
-({\bm \mu}^{(1)} + {\bm \mu}^{(2)})\cdot{\bm B}, \\
\hat H_{\mathrm{I}} & = & \sum_{K=0,2,4} \sqrt{2 K + 1} 
\left[{\bm C}_{K} \otimes {\cal P}{\bm T}_{K}{\cal P}\right]_{0,0}, 
\end{eqnarray}
which is a function of internuclear distance $R$. ${\bm L}$ is the operator of the angular momentum associated with the rotation of the molecular frame, $m_{\mathrm{red}}$ is the reduced mass of the system, ${\bm \mu}^{(1)}$ and ${\bm \mu}^{(2)}$ are magnetic dipole operators of the two atoms, and ${\bm B}$ is the vector of the magnetic field. The term $\hat H_{\mathrm{I}}$ is the effective interatomic interaction Hamiltonian acting in the electronic subspace of two $(nsnp) ^3P_2$ atoms, as indicated by the projection operator ${\cal P}$. $\hat H_{\mathrm{I}}$ is a totally invariant tensor, consisting of three distinct contributions, which derive from coupling of rank-$K$ tensor operators ${\bm C}_{K}$ and ${\bm T}_{K}$. ${\bm C}_{K}$ is associated with molecular rotational motion, ${\bm T}_{K}$ describes specific electronic couplings. Electric dipole-dipole dispersion forces contribute to $K=0, 2,$ and $4$, magnetic dipole-dipole interaction enters ${\bm T}_{2}$, and electric quadrupole-quadrupole interaction contributes to ${\bm T}_{4}$ \cite{santra03}. For the incident channels of interest, and for magnetic fields of a few gauss or stronger, all collision dynamics at cold temperatures takes place at interatomic distances of about 100 Bohr radii.  Thus, using a pure long-range Hamiltonian is justified. In the present calculation the potential $\hat V$ is represented by a matrix in the coupled angular momentum basis and evaluated using Wigner-Racah algebra.

The potential matrix of Eq. \ref{eq:Hele} can be diagonalized at different $R$, giving a set of adiabatic potential curves. At large internuclear distances, the potential curves converge to nine different dissociation limits, corresponding to nine different values ($M_{el}=-4,\dots ,4$) of the total electronic angular momentum projection on the direction of the magnetic field. We assume that the initial relative kinetic energy is low enough ($<100 \mu K$), so that we may limit our treatment to the $s$-wave ($L=0$) initial channel. However, other channels with  $L>0$ are present in our model as possible escape channels. The technique of magnetic trapping is limited to low-field seeking atomic states, i.e., the states with a negative projection of the magnetic atomic moment ${\bm \mu}$ on the direction of the field. We assume that initially, before a collision, the two atoms  have a specific $M_{tot}$ projection of the total angular momentum including molecular frame rotation. The states with different $M_{tot}$ projections are not coupled and can be treated separately. Thus, $M_{tot}$, unlike $M_{el}$, is conserved during the collision. In the initial channel $L=0$, thus  $M_{el}=M_{tot}$ for this channel.

\begin{figure}[h]
\includegraphics[width=8cm]{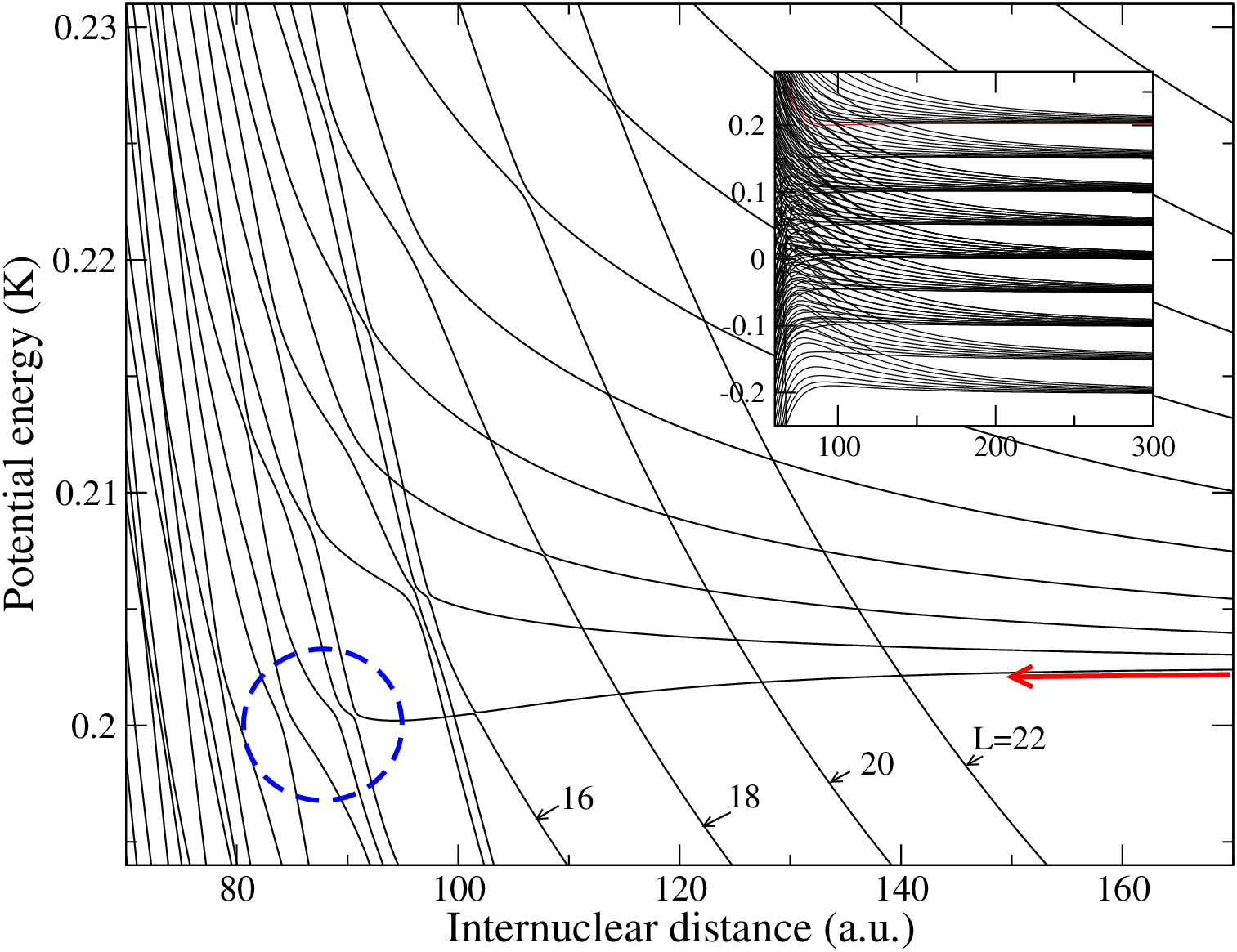}
\caption{\label{fig:ad_curves_art}
Adiabatic potential curves of Sr$_2$[$^3P_2$] calculated for a magnetic field $B=500$ G.  The maximum $L_{max}$  partial wave taken into account for these curves is 22 . The incident channel $(M_{el}=4,\ L=0)$ is marked by a big arrow. The avoided crossing, marked by a circle, produces a large inelastic rate. Numbers at the bottom of the plot specify the angular momentum $L$ of different partial wave channels converging to the $M_{el}=3$ Zeeman level. Note, that at distances of about 100 a.u. different partial waves are mixed. The inset shows the same potential curves on a larger energy scale. In the inset one can see all nine dissociative limits corresponding to nine Zeeman levels of the two separated strontium atoms.}
\end{figure}

Figure \ref{fig:ad_curves_art} gives an example of the adiabatic potential curves for $M_{tot}=4$ at the magnetic field $B=500$ gauss. In the present calculation we have included partial waves of rotating  Sr$_2$ molecular frame up to $L=22$. The entrance $s$-wave channel is marked by a thick line. This channel has a potential well situated around $R=90$ a.u. The well is formed by the interplay of Zeeman, quadrupole-quadrupole and dispersion interactions between the two atoms \cite{derevianko03,santra03}. The shape and depth of the well changes with the magnetic field. This fact allows to change scattering properties of two-body collisions in presence of the field. References \cite{derevianko03,santra03} suggest that at some magnetic field the elastic scattering length can be made very large, allowing fast evaporative cooling.  

The key feature in Fig. \ref{fig:ad_curves_art} is a pronounced avoided crossing at  80-90 a.u. This avoided crossing is mostly formed between the entrance-channel curve and a curve converging to a lower dissociation limit with $(M_{el}=3,\ L=14)$ (see the encircled region in Fig. \ref{fig:ad_curves_art}). Other states are mixed as well in this region, but they play a secondary role. According to the familiar Landau-Zener model \cite{nikitin84}, such an avoided crossing produces a large transition probability from the entrance channel into the $(M_{tot}=3,\ L=14)$ channel.  This is because the energy of the avoided crossing is close to the total energy of the system, and the two curves are almost parallel near the avoided crossing. Thus, at low collision energies, one would expect the inelastic cross-sections to be relatively high.  

In the previous theoretical study \cite{derevianko03}, Derevianko {\it et al.}  have calculated the elastic collisional rates and estimated the inelastic rates as well. Their conclusion was that inelastic collisions do not significantly contribute to the total rate. However, in that study, several simplifying assumptions were made. The most important is that the authors estimated the inelastic losses using a two-channel model. The model includes an $s$-wave channel coupled to a $g$-wave channel by the quadrupole-quadrupole interaction.  Our calculations show, however, that higher, $L>6$, partial waves are particularly important in the loss process. We have found that $L=4$ does not play any important role in the losses, which is in agreement with  Ref. \cite{derevianko03}. This effect of higher $L$ is apparent in Fig. \ref{fig:ad_curves_art}, where the channels  ($M_{el}=4,\ L=0$)  and ($M'_{el}=3,\ L'=14$) are strongly mixed around $R=85$ a.u. At different magnetic field, different partial waves play a dominant role in the inelastic collisions. In other words, the avoided crossing in Fig. \ref{fig:ad_curves_art} can be formed by the states $(M_{el}=4,\ L=0)$ and $(M'_{el}=3,\ L')$ with  $L'$ depending on the field. The inclusion of the higher partial waves not only produces larger non-adiabatic dynamical effects, but it also modifies the shape of adiabatic curves. This would mean that Derevianko {\it et al.} \cite{derevianko03} might underestimate the inelastic rates by neglecting the channels that 1) modify the entrance adiabatic channel and that 2) control the inelastic losses associated with the strong nonadiabatic coupling. The strong multichannel coupling not only causes the large inelastic losses; it significantly modifies  the elastic scattering. This makes the concept of adiabaticity, adopted in Ref. \cite{derevianko03}, inapplicable to the present case.


\begin{figure}[h]
\includegraphics[width=8cm]{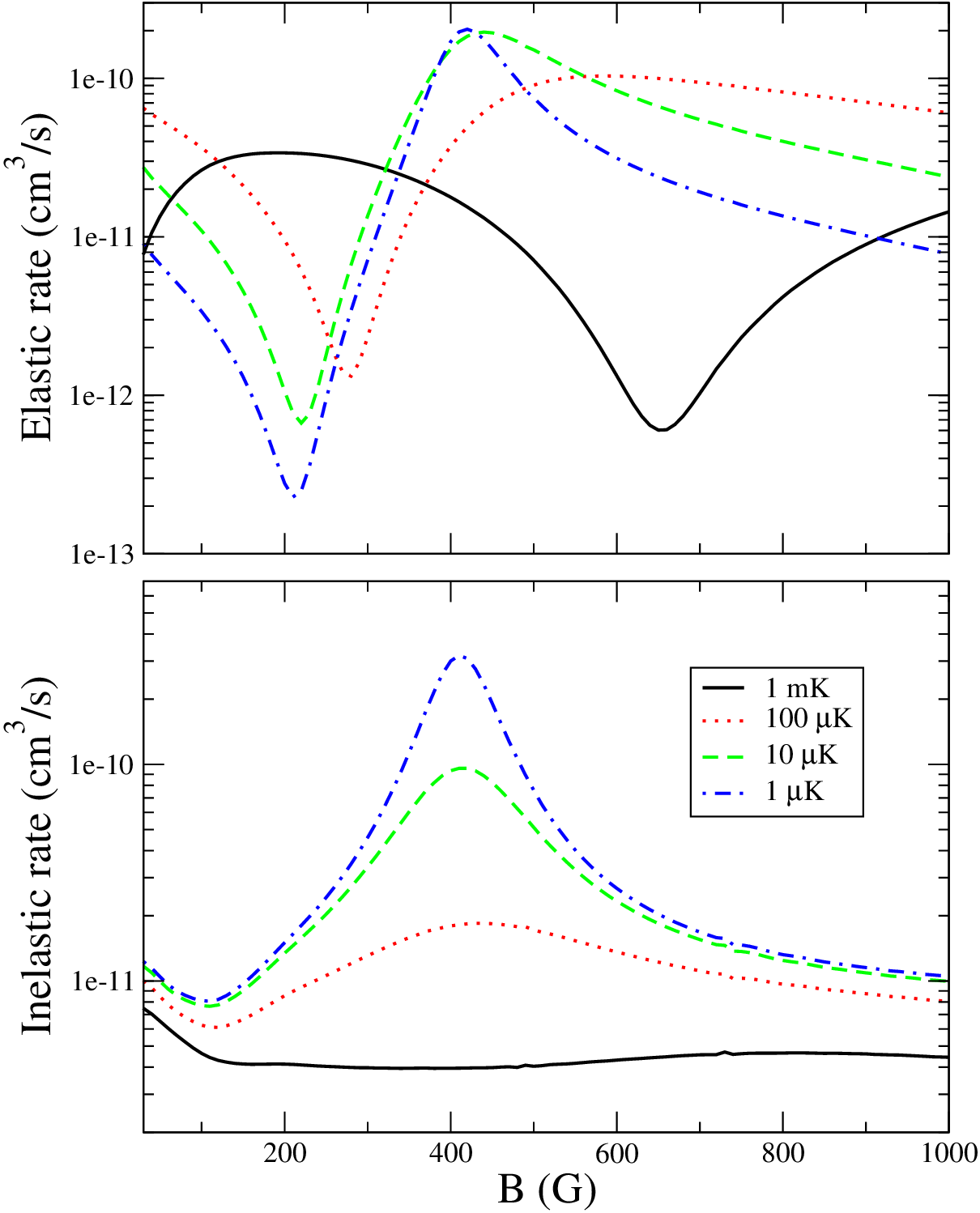}
\caption{\label{Larger_coll_energy_rate_SI}
Elastic (upper panel) and total inelastic (lower panel) rates of the Sr$[(^3P_2)]+$Sr$[(^3P_2)]$ collisions as a function of magnetic field for several incident relative kinetic energies of the two strontium atoms. The entrance channel is $(M_{el}=4, L=0)$. At $1\ m$K, incoming partial waves with $L>0$ can penetrate their respective angular momentum barriers. Their contribution is not included in this plot. (See Fig. \ref{inelastic_rates_E_art})
}
\end{figure}

The above analysis of the adiabatic potential curves develops a qualitative description of two-body collisions. To obtain quantitative results, we solve the coupled radial Schr\"odinger equations for the potential operator of Eq. \ref{eq:Hele}:
\begin{equation}
\label{eq:Sch}
\left[-\frac{1}{2m_{\mathrm{red}}}\frac{d^2}{dR^2}\hat I+\hat V-E\hat I\right]\vec\Psi(R)=0.
\end{equation}
Here $E$ is the total energy of the system, $\vec\Psi(R)$ is the multicomponent radial wavefunction, $\hat I$ is the identity operator. When solving the equation, we work in the coupled channel (diabatic) representation rather than in the adiabatic representation. This choice is justified by the fact that in the diabatic representation, all diagonal and non-diagonal couplings are smooth functions of $R$. Out of the possible diabatic basis sets we adopt the atomic set, which diagonalizes the two-body potential $\hat V$ at $R=\infty$. 

For a given asymptotic collisional energy in the entrance channel, we obtain the total scattering matrix $\underline{S}$ using the one-dimensional $R$-matrix approach \cite{aymar96,burke99}.  The $R$-matrix is obtained solving the Schr\"odinger equation within $R<R_0=1500$ a.u. using finite elements \cite{burke99} as a representation basis. Having obtained the total $\underline{S}$-matrix, the elastic and inelastic rates $\alpha_i$ as function of $E$ are given by familiar formulas (in atomic units):
\begin{equation}
\label{eq:rate}
\alpha_i=\frac{k_{i'}}{m_{\mathrm{red}}}\sigma_{i,i'}=\frac{\pi}{k_{i'}m_{\mathrm{red}}}\left|{S_{i,i'}-\delta_{i,i'}}\right|^2,
\end{equation}
where $k_i' =\sqrt{2m_{\mathrm{red}} (E-E_{i'})}$ is the incident de Broglie wavenumber, $E_{i'}$ is the energy of the incident channel, and $\sigma_{i,i'}$ is the corresponding cross-section. The entrance channel is specified by index $i'$. In our discussion of collision rates, we do not utilize the scattering length, because it is not energy-independent at energies $E>100 \mu K$ for this system. 

In order to check the reliability of the present results, we tested convergence with respect to different numerical and physical parameters of the problem. Two of these, in particular, deserve mention here. The first is a convergence test of the calculated $\underline{S}$-matrix with respect to the position of the inner boundary in the numerical calculations. Since we have neglected the short-range molecular interaction (we do not have the corresponding data), we have to be sure that no flux penetrates into the inner region, i.e. the collisional dynamics is controlled only by long-range potentials. As a result, we have found that for the $(M_{el}=4)$ entrance channel, at low energies below $1 m$K, no flux penetrates into the short-range region. However, the calculation for all the other low-field seeking magnetic states, $(M_{el}=1,2,3)$, shows that there is a substantial probability of penetration, and therefore, the inelastic rate is expected to be large. These entrance channels cannot be treated unless one includes the short-range molecular dynamics. Thus, the following treatment is concentrated on the $(M_{el}=4,\ L=0)$ entrance channel solely. The second test is an assessment of convergence with respect to the maximum number of partial waves that are included in the calculations. As was discussed above, most of the inelastic losses occur due to the avoided crossing in the potential well of the incident channel. At the largest magnetic field considered here, B=1000 G, the avoided crossing is formed by an interaction with the $M_{el}=3,\ L=14$ channel. Performing the full $\underline{S}$-matrix calculation with maximum number of partial waves, $L_{max}$, up to 18, we have found that, indeed, the inelastic and elastic rates are converged at $L_{max}=14$. This test underlines the effect of the molecular frame rotation on the scattering process.

\begin{figure}[h]
\includegraphics[width=8cm]{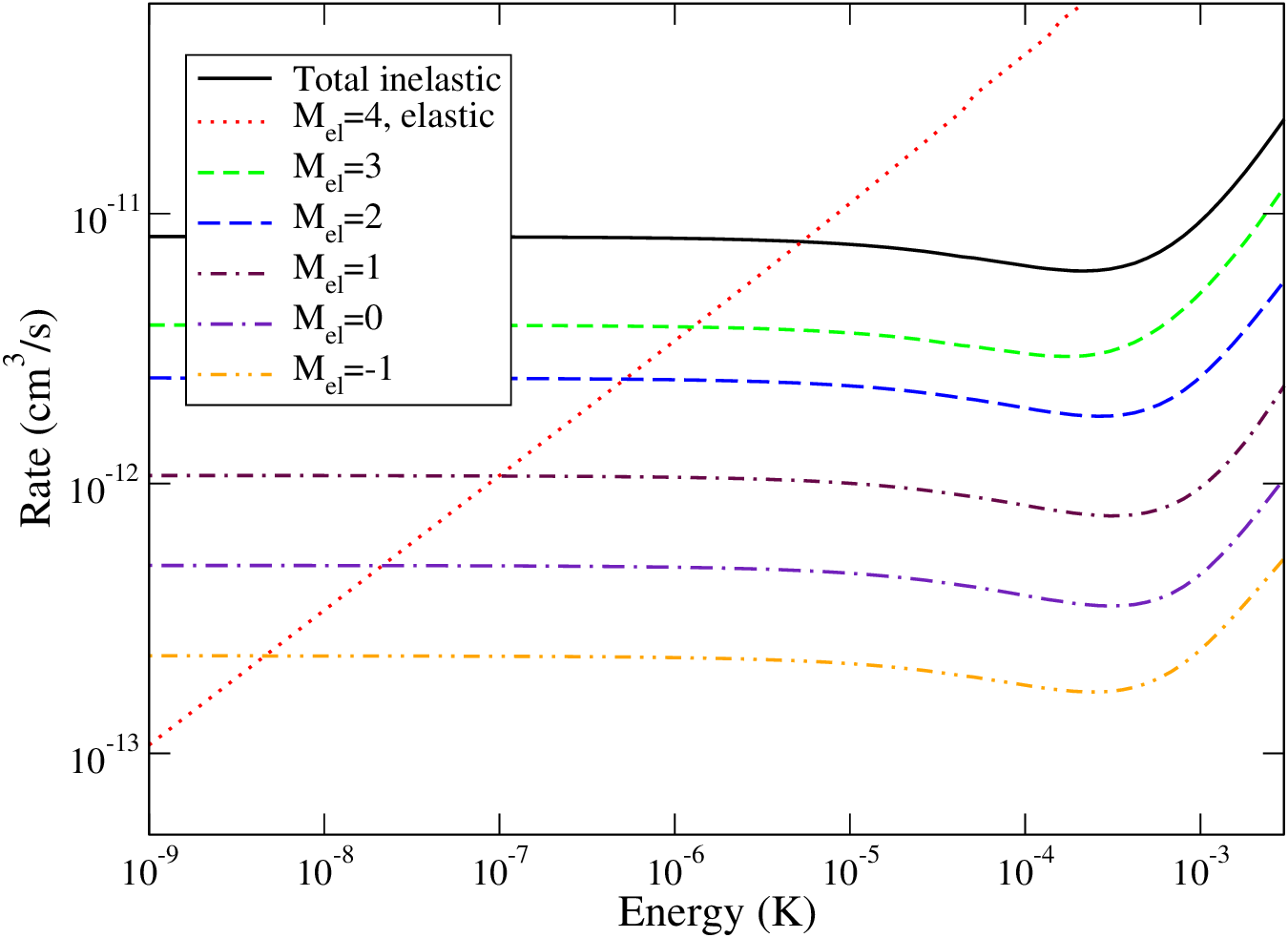}
\caption{\label{inelastic_rates_E_art}
The elastic and inelastic collision rates are shown as functions of the relative incident kinetic energy. This plot separates the contributions to the total inelastic rate from different Zeeman levels. The contribution from the lowest Zeeman levels are not shown since they are very small. For this plot, we have included contributions from all higher $L>0$ partial waves in the entrance channel. The higher partial waves are important at energies above $100\ \mu K$.
}
\end{figure}

Figure \ref{Larger_coll_energy_rate_SI} shows the inelastic and elastic rates as a function of the external magnetic field for several incident kinetic energies. The resonant-like behavior of the curves around 400 G is caused by the emergence of a quasibound level in the entrance channel. Due to the strong inelastic loss, the resonance is significantly washed out. Derevianko {\it et al.} \cite{derevianko03} have predicted a resonance at a field of about $1000$ G. Since they neglected the inelastic losses in their model, their elastic rate diverges to infinity at the resonant $B$-field. In experiments aimed to obtain a Bose-Einstein condensate, one is particularly interested in the ratio between the elastic and inelastic rates.  Figure  \ref{Larger_coll_energy_rate_SI}  suggests that for strontium in the $(5s5p)^3P_2$ state, evaporative cooling will not be efficient at temperatures below $10\ \mu$K.  Figure \ref{inelastic_rates_E_art} shows the rates as a function of the  relative incident kinetic energy.  The rates of inelastic scattering into individual magnetic sublevels are separated. The magnetic field is fixed at $B=100$ G. As was expected from the analysis of adiabatic curves shown in Fig. \ref{fig:ad_curves_art}, the sublevel $M_{el}=3$ is responsible for most of the inelastic losses.

In view of experimental interest in other metastable alkaline-earth atoms, we have carried out a similar calculation for calcium atoms in the $(4s4p)^3P_2$ state and obtained a comparable result for the elastic and inelastic rates. The main difference is that the resonance for Ca$_2$ is shifted toward larger magnetic fields. To assess whether any other species is likely to produce a more favorable elastic/inelastic ratio, we have completed a model calculation in which we artificially change the quadrupole moment, the nuclear mass and the dispersion coefficients (see Refs. \cite{derevianko03,santra03} for definitions of the quantities). Although such a variation of the atomic parameters modifies the results, the main result remains the same, namely an unfavorable elastic/inelastic ratio at submillikelvin collisional energies. We point out that a similar result was recently obtained in the analysis of the collisions between two polar molecules \cite{avdeenkov02} in an external electric field.  Thus, large inelastic rates apparently rule out the possibility of cooling atoms or molecules with large anisotropic multipole interactions down to submicrokelvin temperatures using magnetic traps. 

A possible solution to this problem is to use an optical trap \cite{weber02} instead of a magnetic one. In this context, the situation is similar to the case studied in atomic cesium. Early experiments attempting Bose-Einstein condensate of Cs (see, for example \cite{soding98}) using magnetic trapping approaches had shown that the degenerate regime can not be reached due to  a rapid loss of atoms at short internuclear  distances. In the case of Cs, the losses were caused by a large spin-flip cross-section. Recently, the group of Grimm \cite{weber02} utilized a purely optical trap of Cs atoms and successfully achieved a Cs Bose-Einstein condensate. 

{\it Conclusion.} We have calculated the elastic and inelastic two-body collisional rates of metastable strontium atoms in the [$(5s5p)^3P_2$] state and metastable calcium atoms in the [$(4s4p)^3P_2$] state. At low temperatures, $<1\mu K$, collisions in all low-field seeking states are dominated by inelastic processes. This appears to rule out the possiblity of achieving a Bose-Einstein condensate with these metastable atoms, using magnetic traps.

This work has been supported in part by the DOE Office of Science and by an allocation of NERSC supercomputing resources. R.S. gratefully acknowledges financial support by the Emmy Noether program of the German Research Foundation (DFG).

\end{document}